\documentclass[useAMS,usenatbib]{mn2e}
\usepackage{psfig}

\def\apj{{\rm ApJ}}
\def\mnras{{\rm MNRAS}}

\def\etal{{\rm et~al.\ }}
\def\hmpc{\;h^{-1}{\rm Mpc}}
\def\invhmpc{\;h\;{\rm Mpc}^{-1}}

\def\kms{{\rm \;km\;s^{-1}}}

\def\simlt{\lower.5ex\hbox{$\; \buildrel < \over \sim \;$}}
\def\simgt{\lower.5ex\hbox{$\; \buildrel > \over \sim \;$}}

\title[Peaks in the cosmological density]{Peaks in the cosmological
density field: parameter constraints from 2dF Galaxy Redshift Survey data}

\author[S. De and R.A.C. Croft]{
Soma De$^{3,1}$\thanks{E-mail: somad@nhn.ou.edu} and
Rupert A.C. Croft$^{1,2}$\\
$^{1}$Dept.   of  Physics, Carnegie   Mellon  University,
Pittsburgh, PA 15213, USA\\ 
$^{2}$ Bruce and Astrid McWilliams Center for Cosmology,
Carnegie   Mellon  University,
Pittsburgh, PA 15213, USA\\ 
$^{3}$ Homer L. Dodge 
Department of Physics and Astronomy, University of Oklahoma, Norman,
Oklahoma 73019, USA\\
}
\begin{document}

\pagerange{\pageref{firstpage}--\pageref{lastpage}} \pubyear{2005}

\maketitle

\label{firstpage}

\begin{abstract}
We use the number density of peaks in the smoothed 
cosmological density field taken from the 2dF Galaxy Redshift Survey to 
constrain parameters related to the power spectrum of mass fluctuations,
 $ n $ (the spectral index),
d$n$/dln$k$ (rolling in the spectral index), and the neutrino mass,
$m_{\nu}$. 
In a companion paper we use N-body simulations to 
study how the peak density
responds to changes in the power spectrum, the presence of redshift 
distortions and the relationship between galaxies and dark matter halos.
In the present paper we make measurements of the 
peak density  from 
2dF Galaxy Redshift Survey data, for a range of smoothing filter scales
from $4-33 \hmpc$. 
We use these measurements to constrain the cosmological parameters,
finding $n=1.36^{+0.75} _{-0.64}$,
$m_{\nu} < 1.76 $eV, $d$n$/dln$k$=-0.012^{+0.192}_{-0.208}$, 
at the 68 \% confidence level,
where $m_{\nu}$ is the total mass of three massive neutrinos. At 95\% 
confidence  we find $m_{\nu}< 2.48$ eV.
These measurements represent an alternative way to constrain cosmological
parameters to the usual direct fits to the galaxy power spectrum, and are
expected to  be relatively insensitive to non-linear clustering 
evolution and galaxy biasing.
\end{abstract}

\section{Introduction}
The space density of peaks in a cosmological density field smoothed with a 
filter is sensitive to the shape of the linear power spectrum of
mass fluctuations, even for filter sizes where the fluctuations are
in the non-linear regime (Croft \& Gazta\~naga 1998).
In De \& Croft (2007), hereafter Paper I 
we explored the sensitivity of the peak density 
to parameters related to the initial power spectrum, 
as well as redshift distortions and variations in the galaxy halo occupation 
distribution (e.g., Berlind \& Weinberg 2002). 
The theory of peaks in a Gaussian density field was set out
in detail by Bardeen \etal
(1986, hereafter BBKS). The relationship between the peak
density and power spectrum in BBKS can be used 
to explore  such parameters as the spectral index and its dependence on 
scale (Kosowsky and Turner, 1991) along with the neutrino mass. In this 
paper we use data from the 2dF Galaxy Redshift Survey (hereafter
2dFGRS, Colless \etal 2001),
to constrain cosmological parameters using the peak density.

In a recent analysis of the 2dFGRS 
final data set Cole \etal (2005)  employed a direct Fourier method 
to compute the power spectrum. These authors
put constraints on several parameters using the
directly measured power spectrum shape. They assumed a 
primordial form for P(k) with $n=1$ and $h=0.72$
along with a negligible neutrino mass. This gave 
preferred values of $\Omega _{m}h=0.168 \pm 0.016$ and a 
baryon fraction of $\frac{\Omega_{b}}{\Omega_{m}}=0
.185 \pm 0.046$( 1$\sigma$ errors).
This analysis therefore implies a significantly lower mass density 
compared to the $\Omega_{m}=0.3$ often taken as standard. In combination with  
CMB data from WMAP (Spergel \etal 2003)
Cole \etal find $\Omega_{m}=0.231 \pm 0.021$
Also on large scales some  evidence was seen by Cole \etal  of the baryon 
oscillations predicted by CDM models. Our present work is complimentary
to this large scale linear theory analysis, with the peak density enabling
constraints to be placed on the linear power spectrum and 
cosmological parameters from data on smaller scales.

 This paper is the second in a series.
In paper I, we  examined the number density of peaks in 
cosmological simulations and
compared to results from linear peak theory 
(Bardeen \etal 1986, hereafter BBKS) 
over a range of density field smoothing filter scales.
This provided knowledge of the length scales for which agreement between 
theory and simulation can be expected. 
The dark matter simulations used to compare to peak theory
showed good agreement for filter scales between 3-30$\hmpc$.
This was assuming that the mean interparticle separation
was smaller than the filter scale (for agreement at better than the
 $5 \%$ level).
In addition to simple tests using simulations, we created galaxy catalogues
using lists of dark matter halos and the Halo Occupation distribution
formalism (Zheng \etal 2005.)
Good correspondance was found between the peak density measured from  the
 galaxies and dark matter for filter scales $>4 \hmpc$.

The peak density can be used to constrain cosmology through
its dependence on the power spectrum of mass fluctuations, $P(k)$.
 In particular
we work with the asymptotic number density of maxima, i.e. the  number density
of peaks of all heights and which is found by BBKS to be:
\begin{equation}
\langle n_{\rm pk} \rangle =
\frac{29-6\surd6}{5^{\frac{3}{2}}2(2\pi)^{2}R_{\ast}^{3}},
\label{npke}
\end{equation}
where $R_{\ast}$ is defined to be the following ratio of moments of
the power spectrum:
\begin{equation}
 R_{\ast}=\sqrt{3}\frac{\sigma_{1}}{\sigma_{2}}.
\end{equation}
Here
\begin{equation}
\sigma_{j}^{2} =\frac{\int k^{2}P(k)k^{2j} dk}{2\pi^{2}}.
\end{equation}
The scale dependence of $P(k)$ is probed by  smoothing the
density field with Gaussian filter with comoving radius
$r_{\rm f}$, i.e. multiplying $P(k)$ by
$e^{-(kr_{\rm f})^{2}}$. Croft \& Gazta\~naga (1998) found that there 
is an simple relationship that holds to high accuracy between the
local slope of the power spectrum $n$ at wavenumber $k$ and the filter
scale $r_{f}\sim0.4 \pi/k$. The peak number density is therefore sensitive
to the power spectrum slope (but not its amplitude) and the
parameters which govern it can be tested.

Paper I  focussed on several such aspects of the power 
spectrum such as the neutrino mass, the possible 
rolling of the spectral 
index $n$, and redshift distortions. It was seen that upon increasing the 
neutrino mass the peak density decreases, as it does when
the rolling of the spectral index is made more negative.
 The effect of redshift distortions was also quantified,
showing that they act to suppress the number density of peaks on small scales.
This information is useful for the present paper which deals 
with the peak density in redshift space measured from an observed
galaxy catalogue.

Our approach in this paper is to first test our recovery of the true
peak density
using mock catalogues derived from simulations that have had the 2dF masks and 
selection function applied to them.
We then apply the same estimation techniques to the 2dF data,
using the recovery from mock data to estimate the reliability
of the method. 
The 2dF peak density as a function of filter scale is then used
to constrain cosmological parameters.

The plan for the paper is as follows:
In  Section 2 we  describe the simulated observations (mock 
catalogues), including the detailed reasoning behind their use.
We give an overview of the 2dFGRS
 and in Appendix A
describe how the mock catalogues were generated.
The latter
includes a description of the selection function, magnitude limits and
geometrical limitations of the survey and how they were 
applied to N-body simulations to produce mock catalogues.
In Section 3 we detail tests of the process of finding peaks in the
 simulations and mock catalogues, including selection of the 
volume limits for a given filter size and how we deal with incomplete
sky coverage. In Section 4 we  describe the calculation of 
the observed peak density from the 2dF survey data using the completeness 
functions calibrated from mock catalogues. We also describe
various sources of uncertainty, and how they propagate into
our evaluation of cosmological 
parameters. In Section 5 we  present the best fit values of $m_{\nu}$, $n$ and
d$n$/dln$k$ including error estimates, using Markov Chain 
Monte Carlo analysis. 
In Section 5 we conclude and discuss possible future work.

\section{Survey catalogues}

The main aim of this paper is the extraction of accurate 
information about particular cosmological parameters, 
the neutrino mass, slope of the power spectrum and 
variation of spectral index with scale (rolling.)
We can make good estimates of these quantities if we  marginalize
over parameters whose values are already well known.
This  
involves use of present day knowledge regarding certain other 
parameters such as the density of matter, cosmological constant,
and baryon density, 
$\Omega_{m}h^{2}$,$\Omega_{\Lambda}h^{2}$, $\Omega_{b}h^{2}$.

We work with the space density of peaks in the smoothed
galaxy density field.  To recover the actual space density of peaks 
we need to address the issues related 
to the limitations and particular nature of the galaxy survey. 
In the 2dF survey data, limitations arise for
several reasons.
For example certain galaxies may not make it into the survey
because they are faint, but the magnitude limit varies over the 
survey's angular extent. Very bright objects affect the detection of those
nearby, and so there are exclusion holes in the survey. The survey 
geometry also has a complicated nature, making it different from
the uniform contiguous volume best suited for finding and quantifying peaks. 
 
To be able to estimate  the space density of  peaks 
in the Universe from
the observations it is therefore important to be able to
compute a completeness 
function for  each sub volume of the survey region.  We do this
by computing the ratio of the number of 
peaks in simulations with full sampling and no boundary effects to
the peak density estimated from mock catalogues covering the same 
simulation volume. The mock catalogues are simulated observations (simulations
combined with all the relevant observational constraints).

When using the mock catalogues, we make the assumption that the
completeness function which we generate using them can be applied to 
the real Universe. This means that we assume that the simulations which 
are the basis of the mock catalogues are a close enough approximation to the
statistical distribution of matter in the real Universe that the completeness
is valid. We show in Section 3.1  that in practice this is not an important
restriction as the completeness is in fact unity (i.e. no peaks are missed) 
for most of the survey volume we actually use.
In Appendix A we describe how we generate the mock catalogues.
As they are created to have the same geometry and limitations as
 the 2dFGRS we first describe the observational dataset. 

\begin{figure*}
\centerline{
\psfig{file=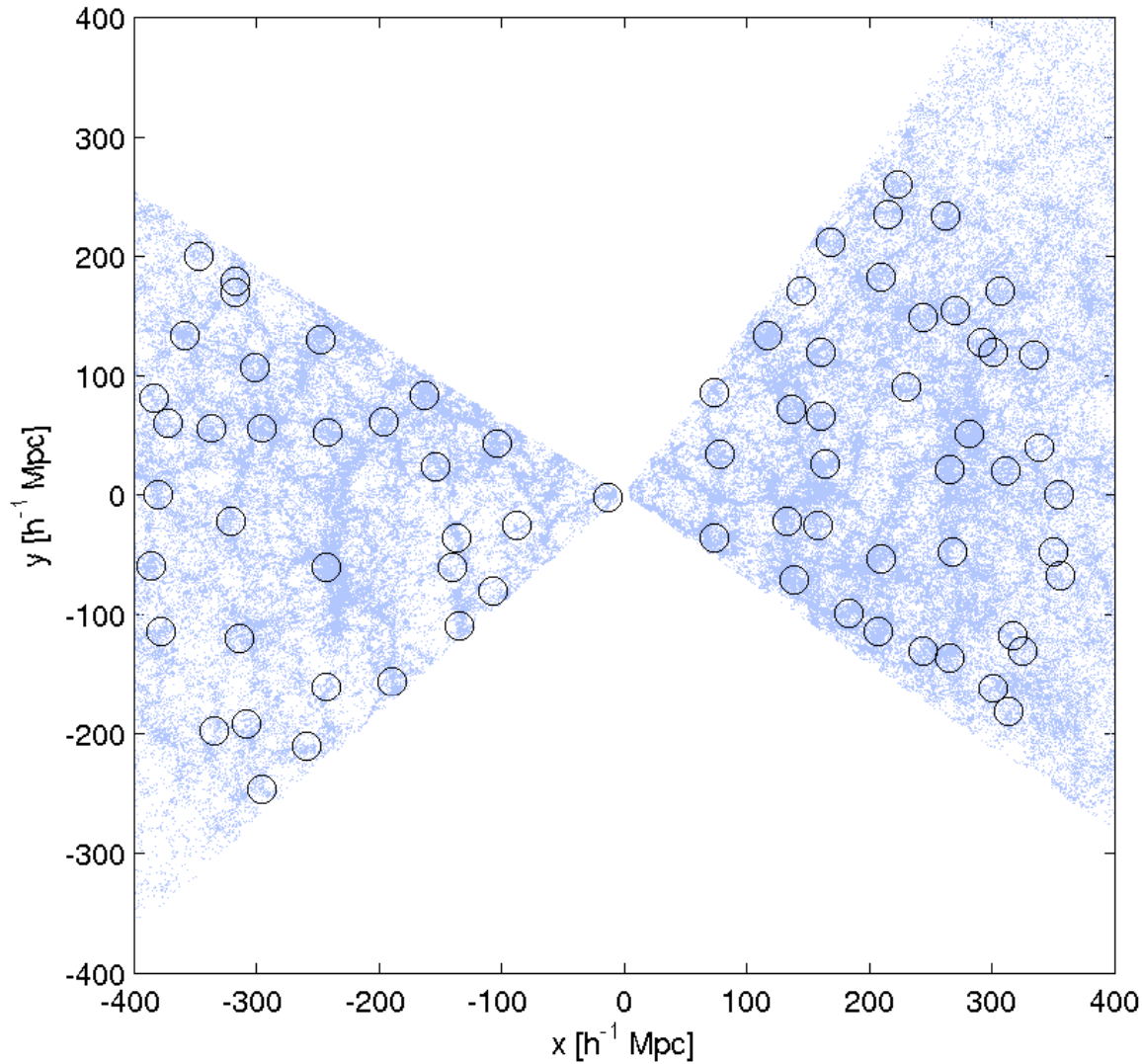,angle=0.,width=17.0truecm}
}
\caption{2D projection onto the Cartesian $x-y$ equatorial plane 
of the 2dF Galaxy Redshift Survey data.
 The points indicate 2dF galaxy positions, and 
the  circles represent peaks in the galaxy density field
when smoothed with a $20 \hmpc$ radius filter. \label{2df}}
\end{figure*}

\subsection{Description of the survey }

    The 2dFGRS (Colless \etal 2001)  resulted in 
 one of the largest catalogues of galaxy redshifts 
 made so far, following a major spectroscopic survey which took 
advantage of the unique capabilities of the 2dF facility at the 
Anglo-Australian Observatory. The 2dFGRS obtained spectra for 
245591 objects, mainly galaxies, brighter 
than a nominal extinction-corrected magnitude limit of $b_{J}=19.45$. 
Reliable (2dF quality flag $>3$) 
redshifts were obtained for 221414 galaxies. The galaxies cover an area of 
approximately 1500 square degrees selected from an extended 
version of the APM Galaxy Survey  (Maddox \etal 1996)
in three regions: an NGP strip, an SGP strip and random fields scattered 
around the SGP strip. In figure \ref{2df} we show a projection of the
galaxy distribution in the 2dFGRS in 
the equatorial plane as well as the positions of peaks found after
smoothing the density field with a Gaussian filter of radius $20 \hmpc$.
(the peak finding will be described later).

Different masks are used when interpreting the 2dF data to characterize the
completeness of the survey in various ways 
as a function of position on sky. The magnitude limit mask gives 
the extinction-corrected magnitude limit of the survey 
at each position on sky. The redshift completeness mask quantifies
the fraction of galaxies above the magnitude limit with measured redshifts.
We use the recommended method  (Colless \etal 2001)
to define this latter mask, it being
specified by the complete set of 2 degree fields that were used to 
tile the survey region for spectroscopic observations. Each point on the 
sky inside the survey boundary is covered by at least one 2 degree field, 
but more often by several overlapping fields. A sector
is defined as 
the region delimited by a unique set of overlapping 2 degree fields. 
This is the most natural way of partitioning the sky, as it takes 
into account 
the geometry imposed by the pattern of 2 degree fields and the way in 
which the galaxies were targeted for spectroscopic observations. Within 
each sector with position angle $\theta,\delta$, the redshift 
completeness $R(\theta,\delta)$, 
is the ratio of the number of galaxies for which redshifts have been obtained, 
$N_{z}(\theta,\delta)$, to the total number of objects contained in the parent 
catalogues, $N_{p}(\theta,\delta)$ so that $R(\theta,\delta) = 
N_{z}(\theta,\delta)/N_{p}(\theta,\delta)$.

We note that when analyzing the data,
the redshift completeness of a given sector, $R(\theta,\delta)$, should be 
clearly distinguished from the redshift completeness of a given field denoted, 
by $C_{F}$, since multiple overlapping fields can contribute 
to a single sector. 
The  $\mu$-mask gives the dependence of the redshift completeness 
on apparent magnitude. We describe the $\mu$-mask in 
more detail in
the Appendix A.
 We use these masks (i.e. values of 
completeness for different patches of sky)
along with the incorporation of redshift limits, magnitude limits 
 and a completeness 
cut-off to make our mock catalogues from simulations (see Appendix A).

\section{Peaks in mock catalogues and simulations}

In order to get a feel for how well peaks that are seen in the fully sampled 
dark matter density field can be recovered in the mock galaxy catalogues,
we carry out a one to one comparison. To find peaks operationally,
we first assign the galaxy (or particle) positions to a grid,
which is then smoothed with a Gaussian filter. Peaks are then
associated with local maxima in this smoothed density field (grid cells
which have a higher density than the surrounding 26 cells).

The first step towards comparing  peak locations in mocks and
the fully sampled 
dark matter density field  is to 
ensure that the same piece of the simulation 
is used in  both the mock catalogues and
the fully sampled case. As mentioned before, our simulations have
 a box size of $300h^{-1}$ Mpc
and we replicate this box, making use of the
periodic boundaries  to get the desired survey volume for the mock catalogues.
We apply the 2dF mask and selection function, resulting in a volume
that matches the shape of the 2df redshift survey.

\begin{figure}
\centerline{
\psfig{file=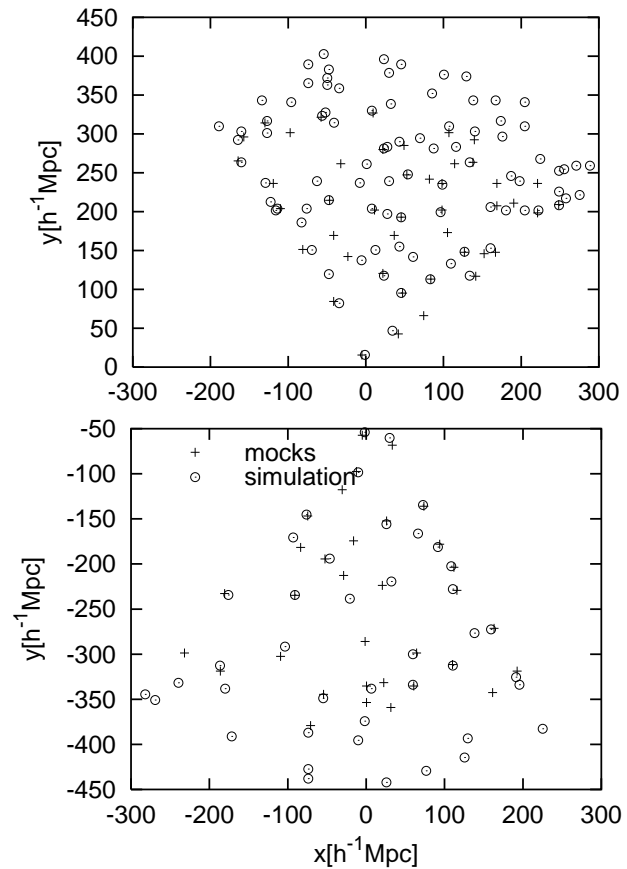,angle=0.,width=8.5truecm}
}
\caption{A test of peak finding in simulated data.
We show a 2D projection of the positions of peaks
found from a fully sampled simulation (circles) and
a mock catalogue (crosses) for a filter scale $r_{f}=10.2 h^{-1}Mpc$.
The top panel shows the SGP and the bottom panel the NGP.
 \label{matching_sim_mock}}
\end{figure}

When finding peaks from the mock catalogues it is important to
assign weights to the galaxies, the appropriate ones being
 related to the inverse
of the selection function. In order to find the weights, we generate
{\em random catalogues}, in which points are initially
 randomly distributed. We apply the 2dF survey mask and selection
function to these points, in the same way as with the mock 
galaxy catalogues. We then
place a grid on top of this distribution of points and count the
number of points in each grid cell. The ratio of the number of points in
the grid cell to the expected number for a uniform distribution gives
the inverse of the appropriate weight to use for that cell. In this way,
 we compute a weight for each cell on the grid. In order to 
make sure the weights have an acceptably low contribution from  Poisson noise
we make the number of points in the random catalogues twenty times
that in the mock galaxy catalogues.

\begin{figure}
\centerline{
\psfig{file=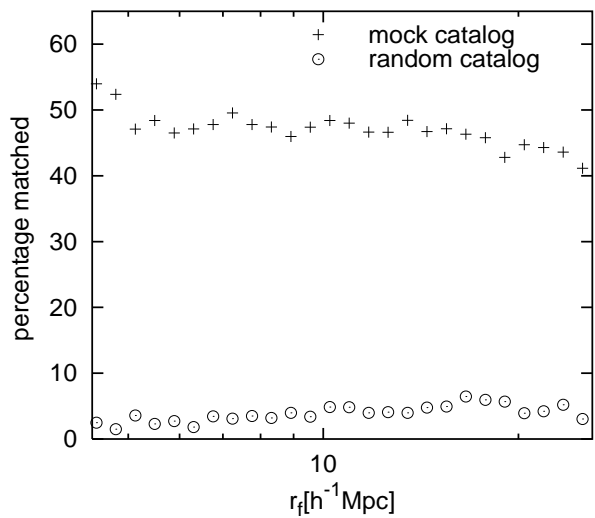,angle=0.,width=8.5truecm}
}
\caption{Percentage of peaks found from mock catalogues (of the NGP region)
within  1 smoothing filter radius of peaks
measured from the fully sampled simulation.
We also show the results for peaks measured from random catalogues.
Both results are shown as a function of filter radius $r_{f}$.
\label{random_match}}
\end{figure}

We   place the same three-dimensional
grid (of size $256^{3}$ cells) onto the distribution of galaxies
in the mock catalogues. The box size was set equal to 
the distance $R_{max}$ as described earlier. We assign the 
weighted galaxy density to the grid 
and smooth this field  in Fourier space with a Gaussian filter. 
We carry this out for several different smoothing
filter radii, $r_{f}$, each time locating the local maxima in the field.

For our one-to-one comparison, we compare directly the positions of peaks
found from the mock catalogues and from the fully sampled simulations.
Because of shot noise, we don't expect peaks in the mocks to be found
in exactly the same places. In Figure 
\ref{matching_sim_mock}, we plot projections of the
NGP and SGP regions for one of the mock catalogues, with symbols showing
where the peaks were found and other symbols showing where the true
peaks lie (measured from the full simulation.) We can see there that the
distrbutions of points are similar in general, but that there are differences,
particularily at the far edges of the survey.

We quantify this in Figure \ref{random_match}
 where we show the fraction of peaks
in the mock catalogues which have a true peak within one filter radius.
We can see that good peak detections, defined in this way occur
approximately 50\% of the time, roughly independent of the filter size.
In the same plot we have carried out this comparison with peaks
found from the random catalogues, to show how likely it is that false peaks
arise from Poisson fluctuations. We can see that this is $\sim10-20$ times less
likely, showing that we are truly extracting peak information
from the mock catalogues.

The one-to-one test carried out in this section is meant to be illustrative
of how well peaks can be identified in galaxy catalogues. 
The information presented in Figures \ref{matching_sim_mock}
and \ref{random_match} is not used in our measurement of the peak
density from observations. The completeness corrections used for this purpose
are described in the next section.

\subsection{Determination of an effective peak completeness}

In order to compare an observed peak density with theoretical models,
we need to statistically correct for the effects of shot noise and 
of survey 
boundaries. Our one-to-one comparison above does not show the aggregate
effect on the peak density of these effects, but only that
approximately 50\% of peaks are found within 1 filter 
radius of their true locations. If we make the assumption that the
2dF observations are affected in the same way by boundaries and shot noise
as our mock catalogues, we can use 
  our mock catalogues to compute an effective completeness (which 
varies spatially).
This will enable us to translate an observed peak density into
the density we would have measured with fully information and
 denser sampling.

\begin{figure}
\centerline{
\psfig{file=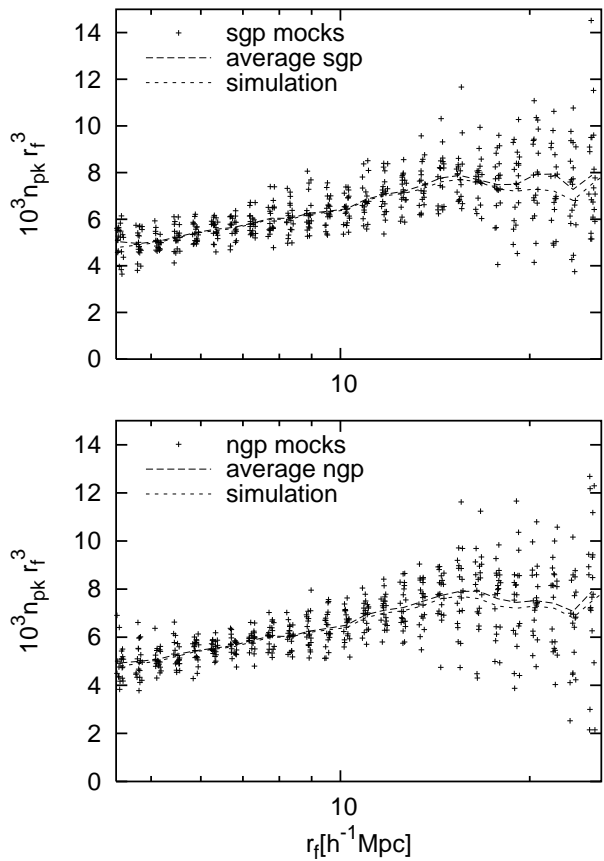,angle=0.,width=8.5truecm}
}
\caption{The peak density measured from the NGP (bottom panel) and SGP
(top panel) mock catalogues.
 The long-dashed lines show the average for 
 the mocks (we plot the corrected peak density when showing the mock
results.) The points are results for each of the 20 mock catalogues.
The short-dashed lines indicate the peak densities measured from 
the fully sampled simulations \label{mock_sgp_ngp}}
\end{figure}   

Our calculations of the effective completeness involve first 
locating the local maxima in the simulation with full sampling
and in the mock catalogues, as described above.
By examining the peak density as a function of distance from the
edge to the survey we have found that boundaries do have a 
substantial effect (for example, in Figure \ref{2df} 
it can be seen that peaks occur preferentially
at the edges) . Rather than applying a large correction for this
effect, we decide to not consider the peaks in the
volume near the boundaries. We choose to remove from our number
density computation all peaks and 
 volume  closer to the side edges of each bin
within $1.5 r_{f}/R_{max}$ in
azimuthal distance and $r_{f}/R_{max}$ in declination 
distance on each side.
Our approach is to only include regions of the simulation
for which the completeness correction
is of order unity (see below.)

\begin{figure}
\centerline{
\psfig{file=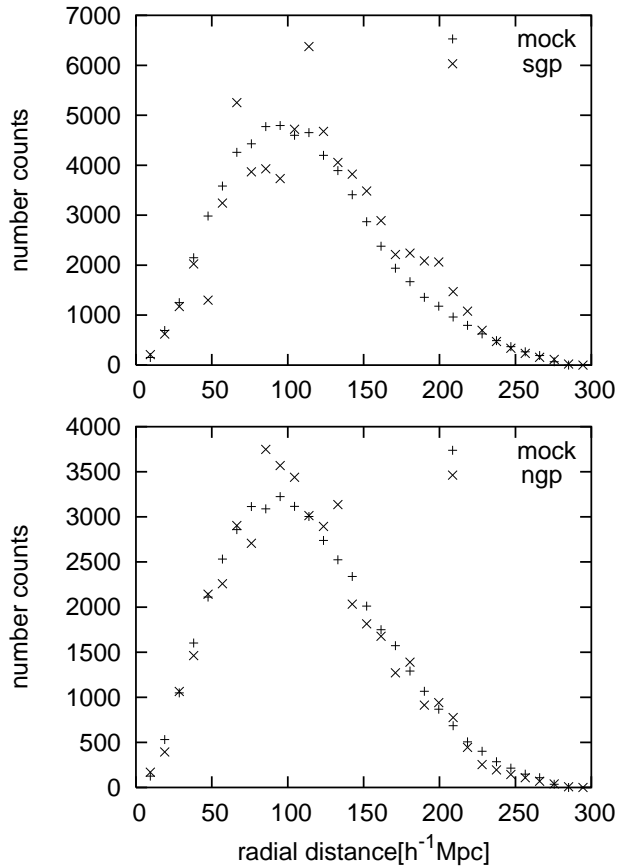,angle=0.,width=8.5truecm}
}
\caption{A
comparison of the redshift distribution of galaxies in one of the mock 
catalogues and the 2dFGRS. We show results for the NGP and SGP regions
separately. The version of the mocks and 2dFGRS used here was that appropriate
for a smoothing filter scale $r_{f}=9.5 \hmpc$ (i.e. the RA and dec cuts
were those appropriate for that $r_{f}$, as descibed in Section 3.3 of
the text.)
\label{number_counts}}
\end{figure}

We compare the number of peaks from fully sampled simulations and
mock catalogues to yield the completeness. To do this we divide space into 
radial bins (distance measured from the origin)
 of width 
$0.5 r_{f}$ for filter radii less than $15h^{-1}$ Mpc and
of width of $0.1 r_{f}$ for larger filter radii.
For each radial bin $i$, a completeness $c_{i}$ given by
\begin{eqnarray}
c_{i} &=& \frac{(n_{sim})_{i}}{(n_{mock})_{i}} ~~~~~~~~~~~~~~~~~~~~~n_{mock}\neq 0 \nonumber\\
c_{i} &=& 0 ~~~~~~~~~~~~~~~~~~~~~~~~~~~~~~~n_{sim,mock}=0 \nonumber\\
c_{i} &=& (c_{i-1}+c_{i+1})/2 ~~~~~~~~~~~~~n_{mock,i}=0,n_{sim}\neq 0\nonumber 
\end{eqnarray}
$(n_{sim})_{i}$ and $(n_{mock})_{i}$ are respectively the number
of peaks in the $i^{th}$ bin measured from the 
simulation and the mock catalogues. The completeness correction
for a particular filter size, $r_{f}$   (denoted 
by $C_{i}(r_{f})$)
is found by averaging over completeness values obtained from 
20 mock catalogues. From an observed number of peaks,
the corrected number of peaks in the $i^{th}$ bin,
  $n_{i}$ can
therefore be computed
from
\begin{equation}
n_{i}=c_{i}(r_{f}) \times n_{obs,i} 
\label{correct}
\end{equation}
where $n_{obs,i}$ is observed number of peaks 
in the $i^{th}$ radial 
bin for the same filter radius. 

For most of the mock survey volume, the completeness we which we compute in
this way is $c_{i}=1.0$,
indicating that we are statistically finding all peaks, with 
no need for any correction. The completeness correction therefore has a
small effect, and there is no significant difference in our results if we
do not include it. The completeness correction does however enable us
to include a larger volume of the survey than without it, resulting in
smaller error bars. The completeness correction $C_{i}(r_{f})$ rises
above 1.0 for large distances from the origin, as the number density
of galaxies goes down. For example, with a filter scale $r_{f}=8 \hmpc$ 
between a radial distance of 300 $\hmpc$ (where  $C_{i}(r_{f})\simeq 1.0$) 
and the edge of the survey volume used
(at $360  \hmpc$) $C_{i}(r_{f})$ averages approximately $1.7$.

In Figure \ref{mock_sgp_ngp}
we have plotted peak densities for different $r_{f}$ values for both 
NGP and SGP mock catalogues. We have also shown with 
 lines the average for mock catalogues. The averaged corrected 
 peak densities (multiplied by $r_{f}^{3}$)
are shown for both the NGP and SGP mocks. We find that for
all $r_{f}$  values they agree very well with the peak densities
 from the fully sampled simulations.

In Figure \ref{number_counts}
we have shown the number of galaxies 
in 30 radial bins for $r_{f}=9.5h^{-1}$ Mpc.
We have already seen that in the very nearby universe it was necessary
to superimpose two mock catalogues in order to achieve
the required number density.

\begin{figure}
\centerline{
\psfig{file=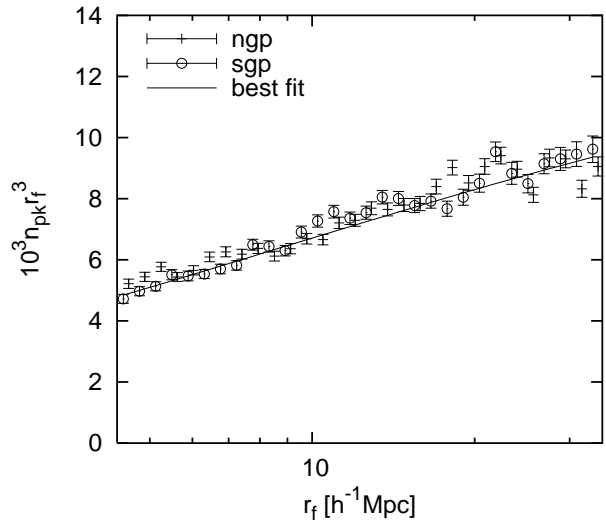,angle=0.,width=8.5truecm}
}
\caption{The peak number density measured from 2dF observations 
as a function of filter scale $r_{f}$.  We show
results for the NGP and SGP regions separately as
symbols. We also show a line correponding to the
best peak theory fit (see Section 5 for details.) \label{obs_peaks}}
\end{figure}

\section{Peak density from the 2dFGRS}

We use Equation \ref{correct} to calculate the peak density in the
observed universe from the 2dFGRS data.
To compute the correction factors,
we use 20 mock catalogues which have a comparable number density of 
objects
to that in the survey. These mocks
 are used to calculate the error bar on each 
data point corresponding to each $r_{f}$. In Figure \ref{obs_peaks} we 
show the peak densities for the SGP and NGP. We also indicate
 the best-fit theoretical curve which
corresponds to a model with $n=0.94$, $d=-dn/dlnk=-0.009$, $m_{\nu}=0.70$ eV
 (total mass of three massive neutrino species in eV).
We descibe how the model fitting was done in Section 5 below.

\begin{figure}
\centerline{
\psfig{file=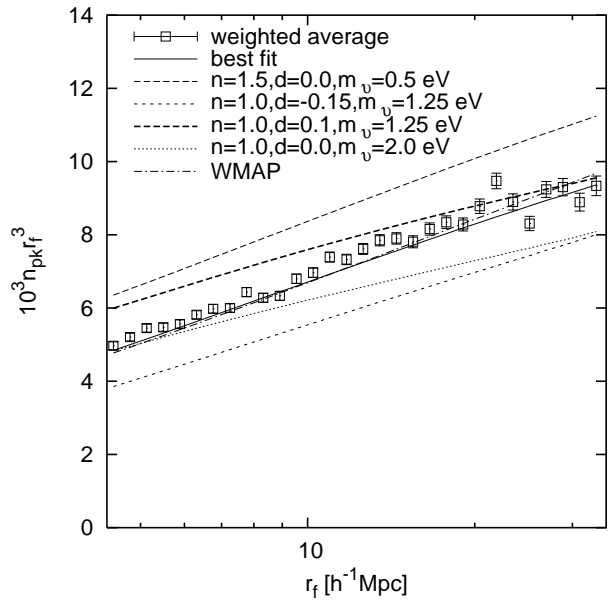,angle=0.,width=8.5truecm}
}
\caption{The observed peak density
measured from the 2dFGRS (points) alongside 
several different peak theory model curves.
The 2dFGRS results are the weighted average of those for the
NGP and SGP regions show in Figure \ref{obs_peaks}. The models
are examples of number densities computed from peak theory
for a few different values of the spectral index, $n$, the running of the
index, $d$ and the total mass of neutrinos, $m_{\nu}$.
 \label{weight_obs}}
\end{figure}

\begin{figure}
\centerline{
\psfig{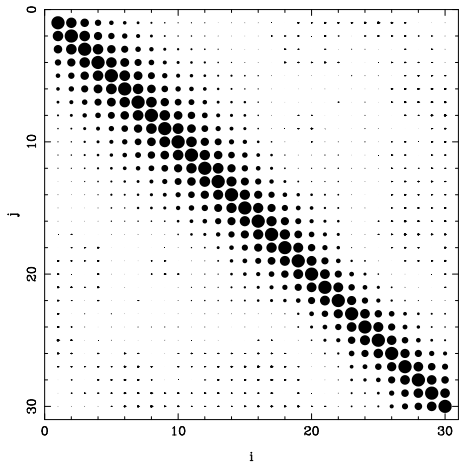}
}
\caption{
The covariance matrix of the peak density $n_{pk}$ computed
using mock catalogs and Gaussian simulations (see \S5.2)).
 The symbol area is proportional to
$C_{ij}/(C_{ii}C_{jj})^{1/2}$, with negative elements shown 
as open symbols. The 30 elements span values of $r_{f}=4.47 \hmpc$ 
to $r_{f}=33.1 \hmpc$, as given in Table \ref{pktab}.
\label{matrix}}
\end{figure}

In Figure \ref{weight_obs} we show the weighted average of the peak
 density measured from the SGP and NGP.
For illustrative purposes, 
we also show  the peak densities predicted by various 
models. The observational weighted average
of NGP and SGP data was calculated using:
\begin{equation}
n_{\rm pk,obs}=\frac{(f_{\rm SGP}) n_{pk,\rm NGP} +(f_{\rm NGP})
n_{\rm pk, SGP}}{f_{\rm NGP}+f_{\rm SGP}},
\end{equation}
where $n_{\rm pk, SGP}$ and $f_{\rm SGP}$ are respectively the 
number density of peaks in the SGP region of the survey and the 
fractional error on that value.

\begin{table}
\centering
\caption[pktab]{\label{pktab}
The space density of peaks measured from the 2dF
redshift survey data, as a function of filter scale $r_{f}$}
\begin{tabular}{ccc}
\hline&\\
 $r_{f}$ & $n_{pk}r_{f}^{3}$ & $\sigma(n_{pk}r_{f}^{3})$ \\ 
 $(\hmpc)$ &  $\times 1000$ &  $\times 1000$ \\
\hline & \\
4.47 & 4.97 & 0.11 \\
4.79 & 5.21 & 0.10 \\
5.13 & 5.45 & 0.11 \\
5.50 & 5.47 & 0.11 \\
5.89 & 5.56 & 0.11 \\
6.31 & 5.81 & 0.11\\
6.76 & 5.98 & 0.11\\
7.24 & 6.00 & 0.11\\
7.76 & 6.43 & 0.12\\
8.32 & 6.28 & 0.12\\
8.91 & 6.33 & 0.12\\
9.55 & 6.80 & 0.13\\
10.2 & 6.97 & 0.13\\
11.0 & 7.39 & 0.14\\
11.7 & 7.33 & 0.14 \\
12.6 & 7.61 & 0.15\\
13.5 & 7.85 & 0.15\\
14.5 & 7.90 & 0.16\\
15.5 & 7.81 & 0.16\\
16.6 & 8.16 & 0.17\\
17.8 & 8.34 & 0.17\\
19.1 & 8.28 & 0.18\\
20.4 & 8.78 & 0.20\\
21.9 & 9.47 & 0.21\\
23.4 & 8.90 & 0.22\\
25.1 & 8.31 & 0.19\\
26.9 & 9.24 & 0.22\\
28.8 & 9.30 & 0.24\\
30.9 & 8.90 & 0.24\\
33.1 & 9.34 & 0.27\\
\end{tabular}

\end{table}

In order to carry out the maximum 
likelihood fitting described in the next section
we compute the error in our measurement of the peak density from 
observations.  We make use of our mock  catalogs to quantify the
different components.

 We include the contribution caused by cosmic variance.
This we estimate from the standard deviation about the mean
in the number of peaks in the different mock catalogues.  
We compute this
for the 25 different filter radii ranging from $4.5 \hmpc$ to $33 \hmpc$,
using the 20 
different mock catalogues. We also use the scatter between
the mock catalogues to quantify the effects of variations in the 
completeness corrections on the final result. For the error associated
with the weighted average of the NGP and SGP regions, we use error
propagation,  finding the result to be very close to
$\frac{1}{2}(\sigma_{ngp} ^{2} +\sigma_{sgp}^{2})^{\frac{1}{2}}$.

\section{Constraints on Cosmological parameters}

\subsection{CDM parameters}

In order to see which values of cosmological parameters are allowed by
the data, we fit the 
peak density we measure from the 2dFGRS observations with that expected in 
different versions of the CDM cosmology. We concentrate on
 three parameters, the spectral index $n$, the running of $n$, 
${\rm d}n/{\rm dln}k$,
and $m_{\nu}$ (the total mass of three massive neutrinos species in eV).
For the other parameters which affect the shape of the power spectrum, we
use values from the WMAP first year data release
(Spergel \etal 2003). These are $\Omega_{m} h^{2}=0.135 ^{+0.008}_{-0.009}$,
$h=0.71^{+0.04}_{-0.03}$, $\Omega_{b}=0.0224\pm0.0009$. Our results are
insensitive to adoption of  the WMAP5 parameters instead (Dunkley \etal 2009).
We have seen in Paper I that  the effect
of small scale redshift distortions can
be parametrized using the one dimensional
velocity dispersion of galaxies, $\sigma_{f}$ . We use a prior on its
value in our analysis:  $\sigma_{f}=3.4 \pm 1.0$ (in units of $100 \kms$),
based on the value given Matsubara \etal (2004). The Gaussian error on the
value we use is large enough to incorporates other recent measurements
at the $1 \sigma$ level such as that of
Peacock \etal (2001) ($400 \kms$). 
We also use a prior on the running of the spectral index,
 $\frac{dn}{dlnk} = 0 \pm 0.3$.

\begin{figure*}
\centerline{
  \psfig{file=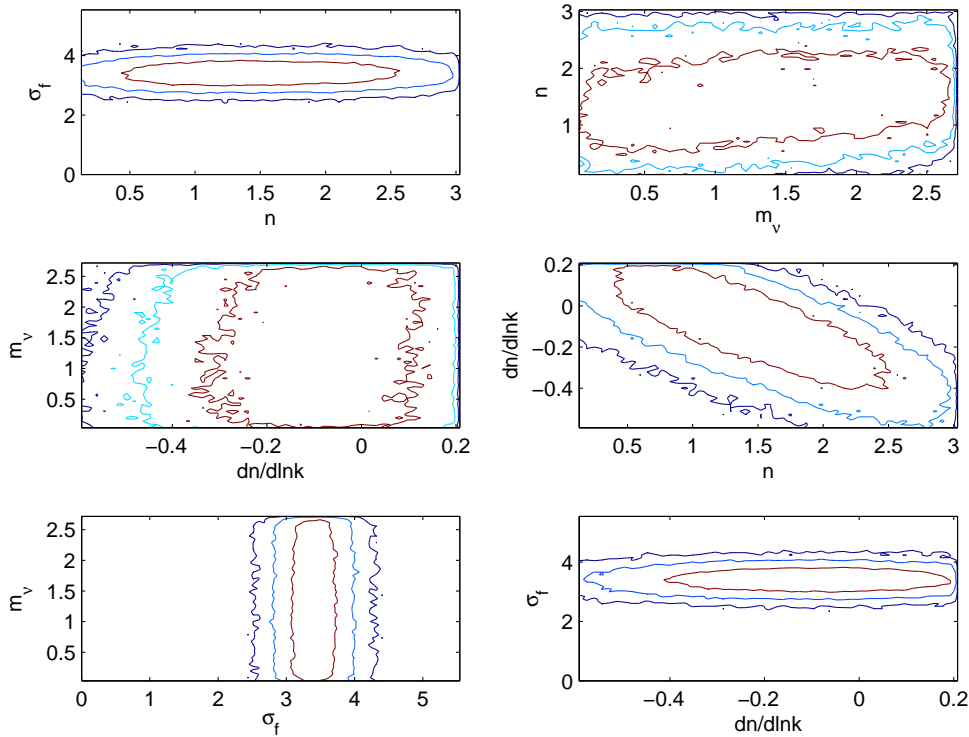,angle=0.,width=15.0truecm}
}
\caption{Probability density contours (at the $1, 2$ and $3 \sigma$ level)
derived from the Monte Carlo Markov Chain applied 
to the 2dFGRS peak data for the joint
confidence interval on pairs of cosmological parameters. We
show results for the spectral index $n$,
the running of the index ${\rm d}n/{\rm dln}k$, 
the neutrino mass $m_{\nu}$ (in eV) and the 1d velocity dispersion
of galaxies, $\sigma_{f}$ (in units of 100 $\kms$)\label{MCMC}}
\end{figure*}

\subsection{Construction of the Covariance Matrix}
In our analysis we compute the likelihood as $\propto e^{-\chi^{2}/2}$,
where  the $\chi^{2}$ values use the full covariance 
matrix to include the correlations between bins at different scales.
We note that we have found that including only the diagonal elements
would give erroneously tighter constraints (at the factor of $\sim 2$ level)
on cosmological parameters. 
 We would like to be 
able to calculate the covariance matrix using our 20 mock catalogues,
but with only 20 of them it is too noisy to invert. We therefore use
the mock catalogs to compute the diagonal elements only.
We then use 1000 
Gaussian realizations of the density field to compute an alternative,
less noisy version of the covariance matrix. We scale the off
diagonal elements of this matrix so that they are consistent with 
those from the mocks (see below).

Operationally, each element of the  matrix is defined from
\begin{equation}
C_{ij}=\sum (x_{i}-\bar{x_{i}})(x_{j}-\bar{x_{j}}) 
\label{cij}
\end{equation}
where $x_{i}$ is the value of $n_{\rm pk}$ for $r_{f}$ bin $i$ and the 
mean $\bar{x_{i}}$ is computed from the 20 mock datasets
or 1000 Gaussian realizations.

We 
compute the $\chi^{2}$ of the difference between observational
measurements and a theory curve using 
\begin{equation}
\chi^{2}=\sum (z_{i,{\rm obs}}-z_{i,{\rm theory}})(C^{-1})_{ij}(z_{j,{\rm
    obs}}-z_{j,{\rm theory}})
\label{chi}
\end{equation}
In  equation \ref{chi} $z_{i,obs}$ is $r_{f}$ bin $i$  from the 
observations and similarly $z_{i,theory}$ indicates the $i$-th  theory 
bin. The sum is over all the $r_{f}$ bins in the dataset (in our
case there are 30 bins). 
In using the covariance matrix the main challenge is the inversion of
 $C_{ij}$ to calculate $\chi^{2}$. We have
chosen the  singular value decomposition 
technique to do this. 
We have plotted the elements
of   $C_{ij}$ in graphical form in 
Figure \ref{matrix}.  We can see that the non-zero elements are concentrated
close to the diagonal, as expected, but that there is still significant
covariance between bins which must be accounted for, so that we need to
invert the matrix and use equation \ref{chi}.
Even with 20 mock catalogues,  $C_{ij}$ is quite 
noisy, making inversion difficult. 

To deal with this, as
 stated above, we  make Gaussian linear theory simulations (with 
the same linear theory power spectrum as our N-body simulations,
and the same box size). We compute the peak number density in 
these simulations and use them to
 compute the off-diagonal elements in the covariance matrix, using
 $C_{ij}=(C^{\rm mock}_{ii}
C^{\rm mock}_{jj})^{1/2}C^{\rm Gauss.} _{ij}/(C^{\rm
Gauss.}_{ii}C^{\rm Gauss.}_{jj})^{1/2}$.
Here $C^{\rm mock}_{ij}$ are elements computed
from the mocks and $C^{\rm Gauss.}_{jj}$ from the Gaussian simulations.
We do this for each half of the survey.
The resulting matrix elements for the NGP
are then summed with those for the SGP calculated in a similar fashion
and this matrix is the one used to compute 
equation \ref{chi} in our analysis .

\subsection{Markov Chain Monte Carlo}

In order to compute constraints on the multi-dimensional
space of parameters efficiently, we use use a 
Monte Carlo Markov Chain approach. The procedure that we
follow is that of Verde \etal (2003).
As in that paper we 
 generate  Markov chains which have a large number of  
 values (30000) to ensure a satisfactory level of convergence. 

\begin{figure*}
\centerline{
\psfig{file=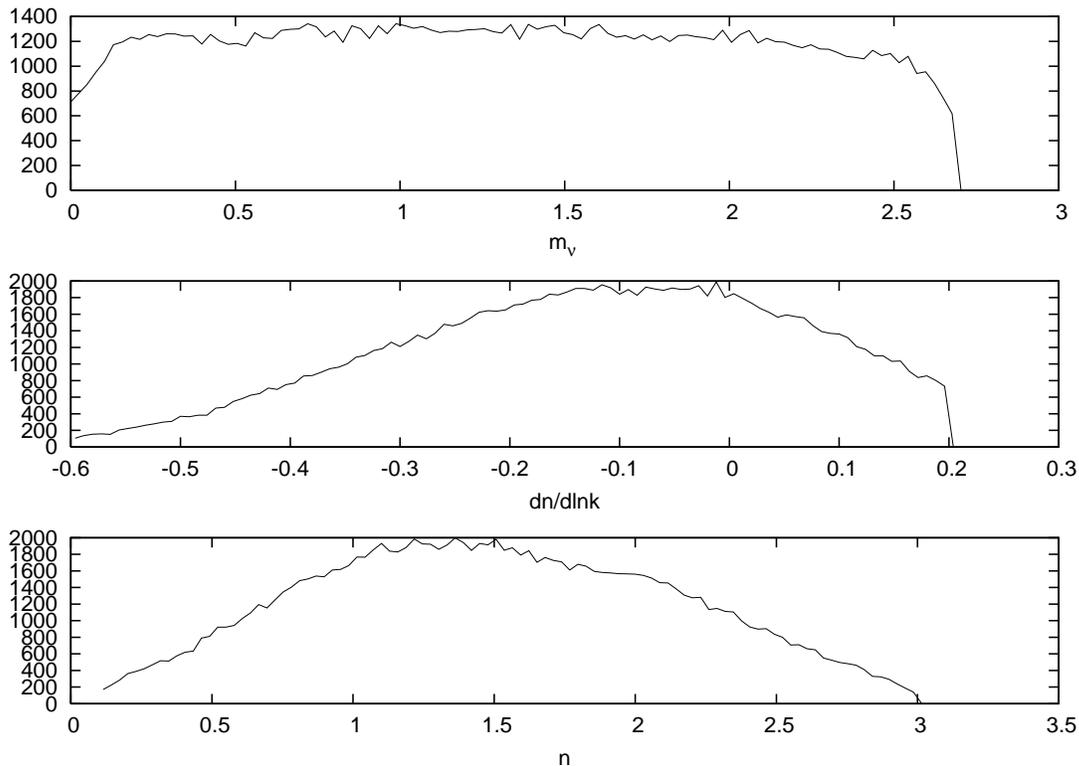,angle=0.,width=15.0truecm}
}
\caption{
One dimensional probability distribution of cosmological parameters
measured from the 2dFGRS data.
We show results for spectral index  $n$,
the running of the index ${\rm d}n/{\rm dln}k$, 
and the neutrino mass $m_{\nu}$ (in eV).
The $y$-axis on each panel gives the number of Monte Carlo Markov Chain 
points in each bin.
 \label{MCMC1d}}
\end{figure*}

To generate 
the Markov chain we compute theoretical peak densities 
for an initial parameter set of $n, {\rm d}n/{\rm dln}k,\sigma_{f},
\Omega_{\nu}$ values chosen randomly (our results are not 
sensitive to the exact range they are drawn from.)
We compute the 
$\chi^{2}$ of the fit to the peak density data from the 2dFGRS  and then
the likelihood, ${\mathcal L}= e^{-\chi^{2}/2}$.
 We then randomly pick another parameter set using 
 a Gaussian distribution centered on the previous
set  (i.e. mean $0$ and variance equivalent to the
 $\sigma^{2}$ associated with each parameter.) If the likelihood 
found from the new set of parameters is 
higher than before, we include this new parameter set in the Markov
 chain and move on, otherwise we compare a randomly
generated number between 
0 and 1 to the ratio of the new and old likelihoods.
If  the number is lower than the ratio we decide to 
include the new set of parameters in the chain and if it is not we  
return to the old set of parameters instead.
As in Verde \etal we start the Markov Chain
from 4 different random locations in parameter space and combine the
results.

In figure \ref{MCMC} we show the two dimensional probability density 
contours for the $1,2,3-\sigma$ (68.3\%, 95.4\% and 99.7\% levels.) 
We can see that there
are some strong degeneracies between the parameters, with for example
the effect of a large $n$ being balanced by a larger $m_{\nu}$. 
Promisingly, the value of $\sigma_{f}$, parametrizing the suppression
of peak density due redshift distortions does not have a strong effect
on  the other parameters. Overall, we
can see that the points with $n=1$,  $dn/dlnk=0$ and  $m_{\nu}=0$
lie inside the 1-2 $\sigma$ contours, indicating that standard assumptions
for these parameters are reasonably consistent with the data.

\begin{table}
\centering
\caption[ptab]{\label{params}
The best fit values for three cosmological parameters measured from
the peak density in the 2DFGRS data (see \S5.2) for details.}
\begin{tabular}{cccc}
\hline&\\
parameter  &    best  &   1 $\sigma$   & 2 $\sigma$   \\
  &    fit  &  range & range  \\
\hline & \\
  $n$     &   1.36   &     $ 0.72\rightarrow2.12$   &    $ 0.14\rightarrow2.61$       \\
d$n$/dln$k$  &-0.01    & $-0.22\rightarrow 0.18$ &  $-0.33\rightarrow0.20$  \\
  $m_{\nu}$ (eV)& 0.27   &  $ 0.0\rightarrow 1.76 $ &  $ 0.0\rightarrow 2.48 $ \\

\end{tabular}

\end{table}

In Figure \ref{MCMC1d} we show the 1 dimensional marginalized probability
distributions for the 3 parameters $n$, ${\rm d}n/{\rm dln}k$, $m_{\nu}$. 
The width of the likelihood distributions
for  $n$, $dn/dlnk$ are large, indicating that the peak density
constrainty will not be competitive with other measures of these
parameters. The  $m_{\nu}$ probability distributions are 
relatively narrow,
 however.
 
The best fit values of the parameters in the marginalized distributions
and their range of uncertainty are give in Table \ref{params}.
We can see that $n$ is consistent within the large
1 $\sigma$ error bars with the WMAP5 results (Dunkley \etal  2009),
 $n=0.963 ^{+0.014}_{-0.015}$. The 2 $\sigma$ upper limit on the
total mass of 3 neutrino species, $m_{\nu}$ is 2.48 eV. This is in the
same range as upper limits from a variety of other cosmological 
probes (see Elgaroy 2007 for a recent review.)

\section{Conclusions}

By measuring the space density of peaks
in the smoothed 2dF galaxy survey density field as a function 
of filter scale, we have constrained the values of
the cosmological power spectrum parameters d$n$/dln$k$, 
$n$ and $\Omega_{\nu}$. We find values that are consistent within 1 $\sigma$
of those measured from  other cosmological observables
(e.g., Dunkley \etal 2009, Seljak \etal 2005), and are in support
of the standard $\Lambda$CDM scenario.
In linear theory, the peak density is not affected by the amplitude of
fluctuations. As simulation tests have shown that this is also true in 
the non-linear regime, for filter scales as small as $3 \hmpc$, this
means that our constraints on cosmological parameters come entirely 
from the shape of the power spectrum. Because of degeneracies in the shape
of the power specrum with cosmological parameters it was necessary to 
assume prior values for some parameters ($\Omega_{m}, \Omega_{b}$ and
$H_{0}$) from the WMAP (Spergel \etal 2003) results in order
to derive our own limits.

Various sources of systematic errors are possible, when measuring
peak statisics,
for example those related to the fact that the peak density in 
a smoothed field is sensitive to boundary effects, which are difficult
to correct for (unlike the case of the correlation function, for example.)
In the present work, we have made conservative cuts, ignoring space
close to the survey boundaries and validated these using mock
catalogs to test both the overall recovery of the peak density and
the recovery of individual peaks. An additional source of error
comes from the fact that redshift distortions tend to merge peaks together
which are separate in real space, as was seen in paper I. We have dealt with 
this effect, by including a power spectrum suppression term
 due to the small scale random velocity dispersion (see paper I for details
and tests). This is the one part of the analysis for which peak finding 
may be most sensitive to the relationship between mass and light. Our tests
using the halo occupation distribution in Paper I have show that this
is unlikely to be a problem at the level of current observational 
uncertainties.

Our constraints have come from an
analysis of  filter lengthscales between $4-30 \hmpc$.
This approach is therefore complementary to much work
in  large scale structure which has been done by 
analyzing the galaxy power spectrum 
on somewhat larger length scales (wavenumber $k< 0.1 \invhmpc$,
corresponding to wavelengths $\sim \pi/k > 30 \hmpc$).
On these larger scales, galaxy fluctuations are assumed
to be linearly related to mass, and direct measurements of the
galaxy power spectrum are used to constrain the mass power spectrum 
and hence cosmological parameters (e.g., Tegmark \etal 2004). 

Recently, it was shown by Sanchez \& Cole (2007) that shapes  of the power
spectra measured from the 2dF and SDSS galaxy surveys
(e.g., Cole \etal 2005, Tegmark \etal 2004)
are not in agreement, due to the $r$-band selected SDSS
galaxies having a stronger scale-dependent bias. This results in 
differences in the cosmological parameters inferred from the
two surveys which are larger than the quoted measurement uncertainties.
From the tests in CG97 and Paper I, we expect that our peak based
measurement to be more robust  both 
to differences in galaxy bias and non-linear
evolution. In the future it will be useful to compute the peak number density
from the SDSS survey data, in order to check consistency. The larger
size of the SDSS final dataset will enable the use of
larger filter scales and reduce the error bars on parameters.

\section*{Acknowledgments}
SD was supported in part by NSF grant AST-0707704 and Department
of Energy Award Number DE-FG02-07ER41517. We thank the referee for useful comments which improved the paper.

{}

\section*{Appendix A: Mock catalogs}

\subsection*{A1 Construction of Mock catalogues}

In order to incorporate observational constraints into  a 
simulation we need to first compute the the selection function.
We choose our luminosity function to have the form of a Schechter (1976)
function (following Erdo\u{g}du \etal 2004) and define

\begin{equation}
\psi(r) = \frac{\int_{L_{\rm min}(r)}^{\infty} \Phi(L)dL}{\int_{L_{0}(\theta,\phi)}^{\infty} \Phi(L)dL}
\end{equation}

 where $\psi(r)$ is the radial selection function at a distance $r$ which is
obtained by integrating over the luminosity function 
 $\Phi(L)$. $L_{\rm min}(r)$ is 
the minimum luminosity  which could be observed at a distance $r$. $L_{0}$
is the lowest measured luminosity in the survey, which is taken as the
luminosity  $L_{\rm min}(r)$ at the comoving distance $r$  of
$5 h^{-1}$ Mpc, equivalent to a lower redshift cutoff for the data sample
we use.

To choose magnitudes for the galaxies in our mock catalogues,
 we numerically solve
 equation $(1)$ for $L_{min}(r)$ using a random value of the
selection function between 0 to 1.
Using this value of the luminosity we estimate
the apparent magnitude at point ($r$,$\theta$,$\delta$),
$\theta$ being the azimuthal angle, $\delta$ being the declination and 
$r$ being the comoving radial distance.
There is a dependence on angular coordinates that 
 comes from the fact that $L_{0}$  is estimated
as the minimum luminosity at a fixed radial distance but its value
varies with $\theta$ and $\delta$. Once we have found the minimum
luminosity at $r$ from the last integral equation, we use following
equation to compute the apparent magnitude ($m$):

\begin{equation}
 L_{min}(r) =[d_{L}(1 +z)]^{2}(1+z)10^{10.0 -0.4 (m-M_{\ast})}
\end{equation}

where $d_{L}$ is the luminosity distance, $z$ is the redshift, 
and $M_{\ast}=-19.7$, equivalent to the absolute $b_{J}$ limit. 
The extra factor of $(1+z)$ is due to  the inclusion of
the $k$ correction.
    
The mock catalogues are made by applying the mask and luminosity function
mentioned above to the outputs of N-body simulations.
To generate the simulation we use the same cosmology as described
in paper I, an LCDM running spectral model with
spectral index $n=1$, ${\rm d}n/{\rm dln}k =-0.020$, 
$\Omega_{m}=0.3$, $\Omega_{\Lambda}=0.7$, and without massive neutrinos.
We use the transfer function 
described in Eisenstein and Hu (1999).
These simulations are identical to those used in  
paper I with respect to box size ($300 h^{-1}Mpc$), 
mass fluctuation amplitude, and other
relevant parameters (particle number $=3.375 \times10^{6}$).

In paper I, we tested the effect on the space density of peaks
of using a galaxy Halo Occupation
Distribution (HOD, e.g., Berlind \& Weinberg 2002) to create the galaxy 
density field rather
than particles.
We found no measurable difference for filter scales $r_{f} \geq 4 \hmpc$ 
and therefore in the present paper, in the interests of having the
maximal space density of objects, we use particles as proxies for
galaxies in our mock catalogues.

    In order to produce mock catalogues, we first apply the geometrical 
constraints (angular coverage) of the survey to our
simulations. There are galaxies in the 2dF survey at redshifts  
as deep as $z=0.3$. To cover the whole of
this observed volume it is necessary to replicate
 the particle distribution of the simulation 
(box size of $300h^{1-}$ Mpc) so that it can cover a sphere of 
radius  $\sim800h^{1}$ Mpc. To do this, we translate
all particles in the same direction with a comoving distance 
equal to the box size (the simulations have periodic
boundary conditions), repeating this 
in three mutually orthogonal directions to yield enough
volume to inscribe the sphere in it. We note that although the
box has been replicated several times, the actual volume of a
single box
is similar to that of the survey, because of the wedge like geometry of 
the survey.
Having carried out this replication,
  we compute the comoving distance from the observer for our
 LCDM cosmology, and compute the redshifts of all the 
particles, including the effect of  peculiar velocities.

After we select the simulation particles which have 
the same angular coverage as in the 
real survey, we reject those which have
 a sector completeness, $C_{F}$ less than $0.2$.
We assign random 
values to the selection function to calculate $L_{min}(r)$ as described
above. This equation is then solved to obtain
apparent magnitudes for each galaxy. We apply magnitude 
limits (described below) to exclude faint objects from the mock 
catalogues. One important point to note is that
in the survey there is also a bright magnitude limit.
It is necessary to ignore very bright galaxies  as they can affect
 observations of neighboring galaxies. The presence of holes in 
the angular mask to exclude them also results in holes in our mock catalogues.
Our chosen values
for the  bright  and faint magnitude limits are
 $b_{J}=-15.0$  and $b_{J}=-19.4$
respectively. Galaxies falling outside this range are ignored.
Using the computed apparent magnitude we calculate 
 the magnitude dependent incompleteness
 $c_{z}(m,\mu)$, which is given by

\begin{equation}
 c_{z}(m,\mu)=\gamma[1-e^{(m-\mu)}]
\end{equation}

where $\gamma =0.99$ and $\mu$  (different for each mask cell) 
are  parameters which were set by 
Colless (2001) by comparing to a 
simple power law model for galaxy number counts (see equations 
$5-8$ of Colless 2001.) These values of $\mu$ were
obtained from  masks made publically available by the
2dFGRS collaboration \footnote{http://www.mso.anu.edu.au/2dFGRS/}
For each mock galaxy we compare this completeness to a randomly
generated number between 0 and 1, rejecting galaxies
when the number is greater than the completeness.
After this, the final mock catalogue is output. To be able to average over 
random fluctuations we  make  many catalogues from 5
simulations generated with different random seeds
as well as  by randomizing the observer positions in the  simulations.
We use 4 different observer positions per simulation, making 20
different mock catalogues.

One problem we encounter when making the mocks is that
for small distances from the observer the number density
of particles in the simulations 
is less than the number density of galaxies in the survey.
For this region, we clone particles to make up the difference. Our results
are insensitive to whether this is done, as we restrict our analysis
to smoothing filter radii much larger than the particle mean separation
at this distance. We describe this procedure in more
detail below.

\subsection*{A2 Determination of the radial boundary}
\begin{figure}
\centerline{
\psfig{file=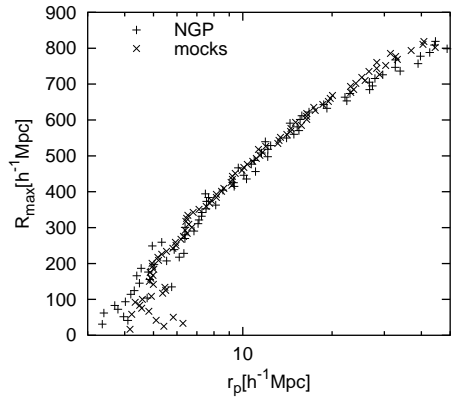}
}
\caption{Variation of the mean intergalaxy separation 
$r_{p}$ in radial bins centered on distance $R_{max}$
from the origin in mock catalogues and 2dFGRS observational
data (for the NGP half of the survey) \label{mean_sep}}
\end{figure}

Before determining the peak density from the catalogues it
is necessary to choose the region  which can
be analysed for each given filter size.  This is because the space
density of galaxies decreases with distance from the origin. If the 
space density of galaxies is too low for a given filter size,
the smoothed density field from the catalogue
will not give a correct representation
of the smoothed underlying density field. 
In paper I, it was shown that the smoothing filter scale
should be greater than the mean separation between galaxies to
avoid serious underestimation of the peak density. 

Additionally, we have a finite sized ($256^{3}$ cells) 
grid which we use to locate
peaks (local maxima on the grid). We make sure at all times that the grid
cell size is also less than the mean separation between galaxies, in
order to avoid missing peaks.
 If {\em w} is the cell width and $r_{g}$ is the mean
intergalaxy distance, then we ensure that $r_{g} \ge w$. 
We have checked to see that this criterion
is sufficient, and our results are not sensitive to this
exact choice.

For a given value of $R_{\rm max}$, we construct a volume limited
catalogue, and compute the mean separation between galaxies.
In Figure \ref{mean_sep} we show the dependence of this average 
inter-galaxy separation $r_{g}$ 
($\sim [V/N_{g}]^{\frac{1}{3}}$, where 
{\em V} is the volume in a radial bin
centered on $R_{max}$ and $N_{g}$ is the number of
 galaxies) 
with distance from the origin $R_{max}$ in both mock catalogues 
and the actual 2dF survey.

As mentioned above, for each filter radius $r_{f}$ we can safely
increase $R_{max}$ to the maximum value where $r_{f} \ge r_{g}$.
In order to determine this value, we 
averaged over the mock catalogues and fitted a fourth order polynomial in 
$\log(r_{f})$ to yield a fitting  function for
the radial distance,$R_{max}(r_{f})$.

\end{document}